\documentclass[12pt]{article}
\usepackage{geometry}              
\usepackage{graphicx}
\usepackage{amssymb}
\usepackage{epstopdf}
\def\fro{\kern-2pt\leftarrow\kern-2pt}

\def\Fi{\8{\imath}}
\def\Vee{{\bigvee}}
\def\0#1{{\mathrm{#1}}}
\def\1#1{{\mathbb{#1}}}
\def\2#1{{\mathbf{#1}}}
\def\3#1{{\mathcal {#1}}}
\def\4#1{{\mathsf{#1}}}
\def\5#1{{\wt{#1}}}
\def\6#1{\overline{#1}}
\def\7#1{{\check{#1}}}
\def\8#1{{\widehat{#1}}}
\def\dag{\dagger}
\def\adj{{^{\dag}}}
\def\dual{{^{\4D}}}
\def\wt{\widetilde}

\def\Bar{\,{\vrule height 10pt width .7pt depth 4pt}\;}

\def\<{\langle}

\def\>{\rangle}

\def\Cliff{\mathop{{\mathrm {Cliff}}}\nolimits} 
\def\Dim{\mathop{{\mathrm {Dim}}}\nolimits} 
\def\Dup{\mathop{{\mathrm {Dup}}}\nolimits} 
\def\Fermi{\mathop{{\mathrm {Fermi}}}\nolimits} 
\def\Grass{\mathop{{\mathrm {Grass}}}\nolimits} 
\def\Min{\mathop{{\mathrm {Min}}}\nolimits} 
\def\Spin{\mathop{{\mathrm {Spin}}}\nolimits} 

\def\ox{\otimes}

\def\x{\times}

\def\BE{\begin{equation}}
\def\EE{\end{equation}}
\def\BEA{\begin{eqnarray}}
\def\EEA{\end{eqnarray}}

\def\lrar{\leftrightarrow}

\def\dar{\kern-3pt\downarrow\kern-3pt}
\def\uar{\kern-3pt\uparrow\kern-3pt}

\newtheorem{definition}{Definition}

\newtheorem{assert}{Assertion}

\newtheorem{assumption}{Assumption}

\newcounter{no}
\begin{document}
\title
{
{\huge Homotopy approach to quantum gravity\\}
}
\author { 
 David Ritz Finkelstein%
\thanks
{ School of Physics, 
Georgia Institute of Technology, Atlanta, Georgia. df4@mail.gatech.edu
}
}
\maketitle
\abstract{
I construct a finite-dimensional quantum theory from  general relativity by a homotopy method. Its quantum history is made up of at least                                                                                                                                                                                                                                                                                                                                                                                                                                                                                                                     two levels of fermionic elements. Its unitary group has the diffeomorphism group as singular limit. Its gravitational metrical form is the algebraic square. Its spinors are multivectors.
}

\section{Strategy}
\label{sec:STRATEGY}

I report here on progress in the search 
for a  physical theory
that covers both  general relativity and quantum theory.
Segal 1951 made a key contribution
 when he pointed out that 
both theories
made a basic classical Lie algebra simple or at least
simpler by a homotopy that  
introduced a small non-commutativity,
with homotopy parameters having as their final values
the physical constants $h$ and $c$. 
He  proposed to carry this to its logical limit, 
a homotopy to a simple Lie algebra,
requiring additional parameters.
Classical physics flattens
this simple Lie algebra.
The strategy of full quantization is to ``flex''  
it back to its natural simple 
form.
Canonical quantization in a nutshell is a homotopy that
``replaces Poisson Brackets by commutators''\/.
Full quantization similarly compressed is a homotopy that
``replaces  commutator Lie algebras by 
simple Lie algebras''.
Einstein and Heisenberg
effected  three homotopies,
with parameters $c$\/, $G$\/,  and $\hbar$.
Full quantization extends all three homotopies
to one that leads to a simple 
Lie algebra.

One may explain the
drift toward  
a simple Lie algebra, and thus 
both relativization and quantization,
in  Darwinian terms.
Within the class of Lie groups, 
the simple Lie groups are stable 
against small changes in the structure tensor,
and the Lie groups of classical mechanics 
and canonical quantum mechanics are compound (not semi-simple)
and are not stable.
In the population of competing theories,
stable ones
probably  survive longer 
and have more offspring than  unstable ones.
Flexing stabilizes, and so promotes survival.

Each epoch defines its own
 stability construct.
For example, Segal stabilized Lie algebras against
variations in  the Lie product but not in
the Jacobi identity or co-commutativity.
I follow the same strategy because it has not yet
been fully executed for any physical theory.

When  we smash a regular algebra flat,  it 
becomes singular and unstable.
One can make it finite
by chopping it into bits of our choosing---
discrete elements ---  and discarding all but a finite number,
but this removes it even further
from nature.
Simple algebras have finite-dimensional representations
giving
all dynamical variables discrete bounded spectra.
Such a quantum theory decides 
its own finite quantum elements.
In some cases, one flexing can quantize
the theory,  relativize it,  stabilize it, and regularize it.
It does not 
preserve the symmetry group of the flattened structure
identically
like renormalization,
nor break it completely
like  lattice regularization,
but flexes it, and so slightly that it continues 
to fit past experiments.

\subsection{Flexing}
\label{sub:FLEX}
Let $\Lambda$ be the manifold of Lie algebra structure tensors on a fixed vector space $V$.
Let $C^n=\{p^k \Bar 0\le p^k\le P^k\}
\subset \1R^n$ be a closed $n$-cell  at the origin 
$\20$ with edges $P^k$\/.
\begin{definition} 
A {\bf flex} $L_0 \fro L_1$ is an $n$-parameter homotopy $h: C^n\to \Lambda$
with unstable $h(\20)=L_0$ ,  
$h(X^k)=L_X$, and
simple $h(p)$  when all $p^k>0$\/.
 \end{definition}
This is an inverse process.  
The direct process is the  contraction, flattening,  or singular limit 
 $\lim_{\20} S = L$ (or $L\to S$ ).
Flexing  a Lie algebra of observables
is a form of  quantization, 
which may be called {\em full quantization}
to distinguish it from canonical quantization,
which is always partial.

Note that people paradoxically call the unbounded  structure a ``contraction'' of the 
bounded, and call the truer structure
 a ``deformation'' of the flattened one.

\subsection{Background}
I came to this problem as follows.
Since 1997 I have taken seriously a
 universal relativity:
that every construct of physics is relative to the experimenter
and changed by its determination.
But every physical theory begins with postulates containing
allegedly absolute
constructs.
Therefore
every physical theory has limited
validity, by the nature of the enterprise.
The goal of physics is therefore not a final theory
but the next step in an
on-going  Darwinian process of theory-selection.
The universe will probably always surprise us.

The search for some theoretical framework 
for this process
sent me back 
to the classic work of Inonu and Wigner 1952 on group contraction,
which now sent me to 
the work of Segal  1951 on the inverse process to group contraction, 
here called flexing.
In his sole illustration, Segal flexed the Heisenberg commutation relations
for one coordinate variable $q$ and momentum $p$ to orthogonal  group relations
\BEA
(qp-pq=i, \quad  iq-qi=0, \quad pi-ip=0)=&&\cr
\quad \lim_{\alpha,\beta\to 0, r\to 1}  (qp-pq=r,\quad rq-qr=\alpha p, \quad pr-rp=\beta q)\/,
\EEA
adjoining and thawing a variable $r$
frozen in the singular limit.

Segal's proposal
influenced the
study of contraction and the Galilean limit by
In\"on\"u and Wigner 1952, and the
 Gerstenhaber 1964  cohomological theory
of Lie algebra stability.
These in turn
had many effects,
many of them first brought out for me 
at this meeting (Oberwohlfach 2006).

Vilela Mendes 1994
first applied
the stabilization strategy to a relativistic physical theory. 
He stabilized 
the Poincar\'e algebra
and the Heisenberg algebra
of Minkowski space, 
using a fundamental length
that could well be the one that Heisenberg 
prophesied 
a half-century earlier. 
His quantum space is  a matrix geometry
 in the sense of Dubois-Violette {\em et al} 1989
except for the theory of connections and gravity.
It is more matricial than  the Matrix Model 
of Banks 1977 in that its time
variable too is  
a matrix,
and it
goes beyond the deformation quantization of Flato 1982
in that its end Lie algebra is simple. 

Vilela space combines and unifies 
 not only the
homotopies
of Einstein
and Heisenberg, but also those of de Sitter, Snyder, and others who navigated
without the stability compass
and so never reached simplicity.
It has the simple group $\0{SO}(6;\sigma)$
 of a quadratic space
 with somewhat unspecified signature $\sigma$.
 
 To apply the stabilization strategy to statistics I define:
\begin{definition} A {\bf paleo-bosonic} statistics
is one defined by a simple Lie algebra having
the bosonic Lie algebra as singular limit.
\end{definition}

Then the  Vilela quantum space 
is an  aggregate 
of paleo-bosonic sub-events that have
a real six-dimensional ket space of 
non-Euclidean signature
(\S\ref{sub:PALEOBOSONS}).
Its event coordinate operators $\8x, \8p, \Fi$
represent elements of  
$d\,\0{SO}(6; \sigma)$\/.
The dimension of the representation 
is unspecified
but is presumably allowed to be large,
since Vilela 1994 uses an infinite-dimensional 
representation for a singular limit.

The dynamics of linear systems like the harmonic oscillator
is expressible by a Lie algebra and is therefore flexible.
Shiri-Garakani and Finkelstein 2005 constructed a stable
dynamics for such a system by flexing the usual unstable one. 
They find that
different time-eigenvalues define subspaces of different dimensionalities, so there is no one-parameter unitary group of time translations, except in the singular limit of classical space-time.
Near the beginning and end of a system time $t$, 
when $|t|\sim\pm l \4X$,
the multiplicities of the eigenvalues of $|t|$ 
vary    
 linearly in $|t|-\0{max}\,\|t|$ and
unitarity is a bad approximation.
This makes room for a quantum version of the black hole.
In the middle times, $|t|\ll \mathrm{max}\|t|$, 
unitarity is a reasonable approximation.

Baugh  2004  flexed the Poincar\"e group
to an $A$ group,
independently of the work of Vilela Mendes with a $D$ group.
The quantum event of 
Baugh space is an aggregate
of  paleo-bosonic sub-events 
with a six-dimensional complex ket space.
For possible future needs
 I generalize Baugh space from  dimension 6
to an extended event space 
of any dimensionality $\nu$
and  signature $\sigma$.
The event coordinate operators $\8x, \8p, \Fi$
then represent elements of  
$d\,\0{SU}(\nu;\sigma)$.

 \subsection{Outline}
In \S\ref{sec:GQ} 
I extend full quantization to the history Lie algebra,
as required for general relativity,
formulating a strategy of general quantization.
In \S\ref{sec:GQC}
I generally quantize the Einstein space, algebra, and group.
In \S\ref{sec:GQK} I generally quantize the Einstein
kinematics and discuss the spin-statistics correlation.
In \S\ref{sec:GQD}
I describe a general  quantum gravity.

I stop in the middle of the work. 
It remains to reconstruct
classical general relativity as singular limit of this quantum theory,
verifying the heuristic arguments used to construct the theory,
and to work out the experimental consequences
that differ from the classical theory.
The main indication that the theory worked would be 
getting some of the particle spectrum and  forces right.
The immediate problem is to get any particles and forces at all,
for they live one or two levels above
the level of events.

\section{General quantization}
\label{sec:GQ}  

\subsection{Foreground} 

 Heisenberg,
emulating Einstein,
set out to work solely with observables,
and ultimately his quantum theory 
encodes operations of observation
in single-time operators $Q(0)$.
But Heisenberg's  dynamical equations 
$dQ(t)/dt=[H(t),Q(t)]$ do not relate 
these alleged observables,
they relate observable-valued functions of time $Q(t)$,
histories.
A history of an observable is not an observable
in the Heisenberg sense.
When he works with admittedly non-observable entities
Heisenberg does not yet strictly
conform to his own operational principle. 
In a regular theory we can come closer  to operationality
with the following
strategy: 
\begin{quote}
\begin{center} {\em Attach observables  to histories,  not instants.}
\end{center}\end{quote}
To formulate quantum dynamics  in classical space-time
in a covariant way,
Dirac, Schwinger, and Feynman already
replaced probability amplitudes for
the instantaneous system 
by probability 
amplitudes for  system histories.
I 
do the same for quantum dynamics
in a quantum space of events.
Continuum  theories of history
have been singular in the  extreme,
with operations
like ``integrating over all histories"
that were  programmatic rather  than  defined 
mathematical processes.
Flexing, however,  makes the quantum history 
more  docile than a
quantum crystal. 
In the fully quantum
theory,
 sums over all histories  are merely traces of  
 finite-dimensional  matrices, while for crystals today
the analogous matrices are infinite-dimensional
 and almost all the traces diverge.
 
To form a general quantum theory I
construct a classical 
history Lie algebra
(\S\ref{sub:UNIDYNAMICS}),
flex it
(\S\ref{sub:UNIPRODUCT}),
and represent it 
in a Fermi algebra (\S\ref{sub:FERMIONIC}).
These steps 
have singular correspondents in canonical quantization,
which is an initial stage of
general quantization that does not yet require us to append
frozen variables.
Instead of a Fermi algebra representation,
the Heisenberg Lie algebra 
uses the 
infinite-dimensional representation 
$R_{\hbar}$ fixed by Planck's constant.
Its unitary flex has instead 
an infinite number of finite-dimensional non-faithful irreducible representations $R_J$ fixed by  weights.
The Clifford and Fermi algebras that arise in general quantization define their own unique faithful irreducible representations up to isomorphism.

\subsection{Unified history}
\label{sub:UNIDYNAMICS}
To start with the easiest  fully
quantum relativistic dynamics,
Finkelstein 2005 flexed  a 
real scalar meson field.
The point of this example was to
quantize the 
history Lie algebra
as a unit,
not
the field variables and the space-time variables separately.
I recapitulate some necessary concepts:
\begin{definition}
\label{def:HISTORY}
A  {\bf q/c} system 
is one with quantum variables 
and classical commuting time and possibly space coordinates,
as in canonical quantum theory;
c/c and q/q systems are similarly defined.

A  {\bf c/c history space} is the space of all possible  state functions
$(q(t), p(t))$.

A {\bf c/c   history algebra}  
is the algebra of all scalar functions of  the history.

A {\bf c/c  history Lie algebra} is the trivial commutator algebra
of the history algebra, which is commutative.

In the {\bf q/c  case}, replace real-valued functions $q(t), p(t)$ by operator-valued functions
in the above definitions.

In the {\bf q/q case}, the history space is
a quantum space defined by  its coordinate algebra,
a flex of a q/c or c/c one.
A history ket defines
a fully quantum dynamics 
by
assigning a probability amplitude to any other
history ket.
\end{definition}
To flex classical general relativity 
I  retrace  Einstein's path to gravity
at the full quantum level of resolution,
one or two levels beneath classical space-time,
starting from the equivalence principle and
flexing as I go.

Invariance under the Einstein group,
the principle of general covariance,
incorporates the equivalence principle and
led Einstein to describe gravity with the
chronometric quadratic form $g_{\mu\nu}$.
The Einstein group  is defined by
 singular relations like $[\partial_{\mu}, x^{\nu}]
=\delta^{\nu}_{\mu}$\/, so it
is  an unstable compound group,
ripe for flexing.

Unlike canonical quantization,
flexing is not intended
to preserve general covariance and
the classical Einstein group exactly.
General covariance is classical
and unstable.
Flexing replaces it with a stable quantum correspondent,
general quantum covariance.

\subsection{Unified algebra}
\label{sub:UNIPRODUCT}
Quantum theory has one product while
general relativity has many.
Part of the solution is
inspired by conversations with
Aage Petersen 1970:
\begin{quote}
{\em Quantization merges products.}
\end{quote}
This already happens in canonical quantization,
which merged
(1) the commutative algebraic product
of functions on phase space
and (2) the non-associative
Poisson bracket product of the same functions.
Heisenberg recovered both from the non-commutative 
product of
quantum mechanics
as leading terms in an $\hbar$ power series.

General quantization merges
(1) the commutative inner product $v\cdot w$
and (2) the anti-commutative Lie product $[v,w]_{\0{Lie}}$,
two products 
of space-time vector fields.
These derive from two 
associative products
on the vector fields:
the differential-operator product
and the Clifford product.
I 
merge them into one
Fermi algebra of  general quantum gravity,
which is also
the operator algebra of the quantum system,
and a Clifford algebra.
The history ket space of the system
is then a multivector space for this Fermi algebra,
and
a spinor space for this Clifford algebra.

A further plurality of products in the theory
arises from the plurality of levels.

\subsection{No-field theory}
\label{sub:NONFIELD}

Hilbert varied 
gravitational field variables $g_{\mu\nu}(x)$ without varying coordinates $x=(x^{\kappa}$).
In the resulting Poisson Bracket Lie algebra, 
$g_{\mu\nu}$ commutes with $x^{\kappa}$.
There are no such coordinates in real life.
The lattice of rods and clocks imagined by Einstein
provides such coordinates 
at low resolution but would
obliterate the system
at high resolution.
Our actual physical coordinates $x^{\mu}$
are all based on weak signals,
usually electromagnetic,
 that  carry us information 
about  the intervening gravitational field 
as well as the remote event,
as in the first astronomical observations 
of the solar deflection of star images.
Such coordinates are more
relative than general relativity,
relative to the field as well as to the frame of reference.
Coordinates in the small that commute with  each other and 
the gravitational field
are unnatural in the canonical theory too, 
as Bergmann and Komar 1972 and Bergmann 1979 point out.
Once coordinates fail to commute we can dispense with 
quantum fields, which re-emerge in the singular limit
of classical space-time.
One way to unify fields is to eliminate them all 
together.

As a classical prototype,  
the gravitational field emerges from a no-field
theory that  imbeds space-time in a higher-dimensional flat space. 
In the quantum correspondent, we imbed a quantum space
of actual events within a quantum space of possible events.

Quantum logic originally forced me to avoid the usual field concept.
Quantum logic
has an invariant construct of subset 
but no general construct of {\em functional} 
relation between given quantum variables, 
as discussed in Finkelstein 1969.
In general quantum gravity 
the main variable, the history, 
is  a quantum set of actual events,
a variable subset of a fixed quantum space of possible
events. 
In fully quantum gravitational dynamics, 
gravity
is already present in  
the quantized event coordinates.

The quadratic form of
general quantum gravity is defined
by the Fermi-algebra product,
which is also a Clifford-algebra product.
 If $v=\lim_{\20} \8v$  is  a classical vector field and
 the singular limit of
an operator $\8v$  in the Fermi algebra,
then the value of the gravitational operator-valued quadratic form is
\BEA
g(v)&=&g_{\mu\nu}(x)v^{\mu}(x)v^{\nu}(x)\cr
&=&\lim_{\20}(\8v)^2\/.
\EEA
Clifford algebra
guides our  general 
quantization
much as the Poisson Bracket Lie algebra
guides canonical quantization.

\subsection{Unified statistics}
\label{sub:UNISTATISTICS} 

The quantum  event of Vilela space has
a paleo-bosonic coordinate algebra of $x$ and $\partial_x$,
as though the event  itself is a paleo-bosonic aggregate.
And the space  is cold, 
so most of its constituents could occupy one ket.
If the Vilela Mendes quantization is correct,
it is odd that history does not consist of one ground event that
occurs very many times and many rare events.
 Instead 
history behaves like a fermionic aggregate;
even a crystal (as Newton noted) with  
its transverse waves.
Crystals are stabilized against collapse
by the fermionic statistics of electrons;
I stabilize history  by a similar strategy:
\begin{quote}
\begin{center} {\em 
Analyze paleo-bosonic events into fermionic ones.}
\end{center}\end{quote}
(\S\ref{sub:GSS},
\S\ref{sub:FERMIONIC}).
This also permits a formulation of the spin-statistics correlation
that makes it natural to extend it to other levels.
The model studied here
is fermionic to its bottom,
several levels of quantification down.

Standard quantum field theory works with 
at least the following successive levels of aggregation,
listed from the top down:
\begin{enumerate}
\item  the many-quantum 
 or field operator history $\7{\psi}(x)$,
 \item the single-quantum ket $\psi(x)$,
\item the coordinate  $x=\int dx$,
\item the differential  $dx$.
\end{enumerate}
Quite different bridges connect these levels.
We pass from
4 to 3 by integration, from 3 to 2 by quantization, and 
from 2 to 1  by quantification.
 Each of these levels has its own algebraic structure,
 4 and 3 being classical and
 2 and 1 quantum.
This arrangement seems unphysical, since 
 surely the microworld is  quantum.

For the purely classical field theory of the 19th century
one mode of aggregation, set formation, would have
sufficed to express all aggregates
and bridge between them.
Canonical quantization must us
classical modes of aggregation
on some
levels 
and quantum on others.
General quantization reopens
the possibility of a uniform
 aggregation process or statistics,
now quantum instead of classical.

\begin{definition}[M and $\mu$]
\label{sub:M} 
If $S$ is a  system (classical or quantum)
then 
$\0M S$  designates
an aggregate whose generic element is $S$.
$\0M$ is for  {\bf Many}-, Meta-,
or {\em Menge}.
If $A$ is an aggregate then in general
${\mu}  A$  
designates the generic element or quantum of  $A$;
$\mu$ is for {\bf mini-}, {mero-,} or  {micro-}.
${\mu} ^n $ and $\0M^n $ designate $n$-fold iterates 
of ${\mu} $ and $\0M$.

${\0V S}$ designates the 
{\bf ket space} of  $S$, 
with norm $\|\psi\|=\psi\adj \psi$
defined by a hermitian form $\dag$.

$\0A \, S$ designates the 
{\bf coordinate $\dag$ algebra} of system $S$,
the endomorphism algebra of $\0V S$.

 The multivalued operation $\Min$ accepts an algebra
and returns  
 a {\bf minimal left ideal} of the algebra.
 \end{definition}

The prefix $\0M$ can designate any quantification process
depending on context or labeling.
In a given context $\mu$ and $\0M$ are inverses of each other
(as in the metric 
system).
I designate the history that we analyze by $H$,
and its constituent events  by
$E={\mu} H$.

\begin{assumption}
\label{ass:BFS}
There is one {\bf basic quantifier}  $\0M$
and it is fermionic.
\end{assumption}
This  is a typical assumption of the uniformity
of nature, based on solid ignorance.
The quantum event is still out of reach,
the chronon more so.
We have a chance of describing it correctly only if it 
repeats what we find on the
higher levels.
At least this assumption does not require us to express bosons like
photons in terms of fermions like neutrinos, as de Broglie
proposed.
I propose to
analyze all the gauge fields as I analyze gravity, 
into fermionic quantum events,  not particles.

Segal's  three variables $p, q, r$ generate the algebra   
$d\0{SU}(2)=
d\0{SO}(3)$,  
in both the $A$ and $B$ series.
In higher dimensions, however, 
one must choose between the $A$ series of 
groups $\0{SU}(n)$ on complex vector spaces
and the $B$ and $D$  series of orthogonal groups $\0{SO}$ on real vector spaces. 
This choice must correlate with the choice of statistics.
There are some tentative indications that the relativity group
is in the $A$ series and that there is a
basic Fermi statistics:
\begin{enumerate}
\item
The $i$-invariance of ordinary complex quantum theory
breaks down in the real $B$ and $D$  series
and survives in the $A$ series.
\item  The singular limit that recovers classical mechanics
from quantum mechanics automatically converts the complex theory into a real one.
\item  Four obvious candidates for a basic quantifier are
$\Cliff$, $\Min \Cliff$, $\Fermi$ and $\Min \Fermi$.
Iterating
$\Cliff$ or $\Fermi$  violates the spin-statistics correlation.
\item $\Min\Cliff$  has a fixed point at dimensions 2 and 4, 
blocking analysis into binary elements. 
\item $\Min \Fermi=\Grass =\Vee = \0M$ is singled out 
because it respects the spin-statistics correlation and permits 
analysis
into binary elements.
\item The internal groups of the Standard Model
are all unitary but not all orthogonal groups, thanks to $\0{SU}(3)$.
\item The paradigm of universal quantification theories,
classical finite set theory,
is an iterated
fermionic statistics
over the binary field of scalars 2.
\item
Fermi quantification has a stable group.
\item
Fermi quantification accounts for both Fermi and Bose statistics at once
(\S\ref{sub:BOSONIZATION}),
and also for spin (\S\ref{sub:FERMIONIC}).
\item The Spin Lie algebra of Minkowski space-time is 
the Lie algebra of an orthogonal group,
not a unitary group; but it is a singular limit of a unitary group,
in the same limit that produces classical space-time.
\item Classical gravity is described by a Clifford ring; but
this is a singular limit of a Fermi algebra.
(\S\ref{sub:RING}, \S\ref{sec:GQD}).
\end{enumerate}
The last two indicators originally pointed to the $D$ series but I
have re-aligned  them by {\em ad hoc} assumptions to
secure consistency.

The Kaluza-Klein strategy imbues space-time
with extra internal compact dimensions.
This created the compactification 
problem: What compactifies these dimensions?
Einstein and Mayer 1931 
evaded this problem by adding extra 
components to
the tangent vectors of the c space-time manifold
but no dimensions to the base manifold itself. 
Connes 1994 makes a similar theory 
in a mixed c-q event space,
attaching quantum spinlike variables to a classical manifold.
Assembling history from fermionic quantum events
permits me to emulate them
at a fully quantum level.
One must posit that these quantum events bind 
along some dimensions of their ket space
to form the macroscopic
space-time dimensions,
leaving all other dimensions small, on the
chronon scale,
like  a soap bubble that is
only one molecule thick in one direction
but  macroscopic in the other three space-time directions.

The compactification problem is replaced by the
extension problem:
What makes some dimensions extend to macroscopic sizes?
As with soap bubbles, this is a matter of
the structure of the molecular elements.
I do not reach this problem here.

\subsection{Regularization}
Almost all quadratic forms are regular,
almost all matrices have inverses,
almost all determinants are non-zero.
The  divergent
cases are
 exceptional, rare.
Therefore any singular theory is not
based entirely on experiment,
which is always generic, but also on
belief in some occurrence of zero probability.
Flexing
 eliminates such singularities
as the Wronskian singularity of gauge theories
and the singularities of propagators.
Instead of infinite renormalizations flexing introduces
quantum constants that are up front and finite.
Vilela space has
three new homotopy parameters and quantum constants:
a space quantum $\4X$,
a momentum quantum $\4P$, and a large quantum number
$\4N$,
in addition to the 
usual quantum of action and angular momentum  
$\hbar$\/.

The  scalar meson in Minkowski space-time  and 
general relativity both have  Lie algebras 
which are infinite dimensional
 because their elements depend on arbitrary functions,
 for example functions of time.
General quantization replaces these Lie algebras by ones
 of high but finite dimensionality.

\subsection{System structure}
To structure the history $H$ we must 
analyze aggregates into elements  more than once.
Present quantum field theory uses several 
traditionally distinct aggregation processes: 
\begin{enumerate}
\item{\em Summation} makes a whole that is ``the sum of its parts,''
at least for some properties such as charge.
\item{\em Integration}  generates quantities from their differentials.
\item{\em Quantification} creates an aggregate of elements. 
The logician William Hamilton introduced
``quantification'' 
to transform yes-or-no questions 
about an individual
 into how-many questions 
about an aggregate.
\item{\em Exponentiation}  
creates $F^{X}$ as an aggregate of $F$-systems, one at each point of the set $X$.
\item{\em Group  generation} creates finite transformations from  infinitesimal ones. 
This constructs groups from their infinitesimal Lie algebras.
\item{\em Quantization}  can be regarded as atomization followed by quantification (\S\ref{sub:QUANT}).
\end{enumerate}
In general quantum relativity all these are provisionally replaced by one aggregation
process.

\subsection{Quantization and quantification}
\label{sub:QUANT}
Quantification in the quantum domain,
usually Bose or Fermi,  formally resembles 
quantization 
so closely that at first it was called 
``second quantization'',
although
quantification converts c systems into c, and q systems into q,
while
 quantization converts c systems into q.
 Nevertheless the two are related.
\begin{assert}
\label{ass:QUANTIZATION}
Quantization is quantum atomization 
followed by quantification.
\end{assert}
{\bf Argument}\hspace{5pt}
(heuristic).
Quantization begins by forming a Lie algebra. 
The vector space supporting the Lie algebra 
can be interpreted as the input/output vector space 
of a constituent or ``atom'' of the system.
The relations of the Lie algebra define a quantification
for this atom.
$\square$
\vskip 10pt
For example, to
 canonically quantize the linear harmonic oscillator, 
with coordinate $x$ and momentum $p$, 
one can:

 (1) Posit
a boson ``atom''  $B$
whose sole attribute is existence,
 with one-component complex normalized 
kets $|B\>$
and bras $\<B|$; and then

(2) Quantify $B$,  forming a bosonic aggregate $\0M B$
with two dimensionless Lie-algebra
generators, a $B$-creator $b\adj=\<\beta |B\>$
and a $B$-annihilator $b=\<B|\beta\>$, subject to
\BE
[b\adj ,b]=c
\EE
where $c$ is a third basis element that is central.
Here $\<\beta|$ is an operator-valued form
converting kets into bosonic creation operators.
Finally one chooses a representation. In this example
the representation is fixed by the value it assigns to the
central  invariant $c = i\hbar$.
The Heisenberg relations are satisfied by 
\BE
x=\4X\frac{(b+b\adj)}2, \quad p= \4P\frac{(b-b\adj)}{2i}
\EE
$\4X,\,\4P$ are dimensional constants required 
for homogeneity of units,
with $\4X\4P=:\hbar$\/.
The classical limit is
$\hbar\to 0$\/.

Quantum electrodynamics too has been quantized 
by (1) atomization and (2) quantification.
Akhiezer and Berestetskii 1953  
(1)
reinterpret
Maxwell's Equations as a one-photon Schr\"odinger Equation,
and
(2) quantify this photon with boson statistics.

\subsection{Paleo-bosons}
\label{sub:PALEOBOSONS}
Baugh 2004 flexed  the Heisenberg Lie algebra
into the unitary Lie algebra. 
His homotopy can also be applied to
the bosonic commutation
relations by a change of basis.

Vilela space and Baugh space are related by an
extension of the Weyl unitary trick.
For brevity I write $V\oplus \22$ as $V$\/.
\begin{definition} A {\bf  unitarization} 
(relative to an orthonormal basis $\3B$ in a 
real quadratic space $V$ with bilinear metric form 
$h_{\mu\nu}$)
 is a process that
\begin{enumerate}
\item replaces
  \BE
 (V, h_{\mu\nu}) \to (V\ox\1C, h_{\mu^*\nu}),
 \EE
a complex $\dag$ space  with Hermitian sesquilinear metric form.

\item imbeds $V\subset  V\ox \1C$ as the vectors with  real coordinates in $\3B$.  

\item maps $\0{SO}(V)\to \0{SU}(V\ox \1C)$,
keeping the same matrix elements in the basis $\3B$.

\item replaces
Clifford generators $\gamma_{\alpha}$ obeying
\BE
\{\gamma_{\mu}, \gamma_{\nu}\} = 2h_{\mu\nu}
\EE 
by
Fermi generators $\gamma_{\alpha}, \gamma\adj_{\alpha}$
obeying
\BEA
\label{FERMIGAMMAS}
\{\gamma_{\alpha}\adj, \gamma_{\beta}\} &=& 2h_{\alpha*\beta},\cr
\{\gamma_{\alpha}, \gamma_{\beta}\} &=&  0,\cr
\{\gamma_{\alpha}\adj, \gamma_{\beta}\adj\} &=&  0\/.
\EEA
\end{enumerate}
\end{definition}
The grade-2 elements $\gamma_{\alpha\beta}$ 
are paleo-bosonic, of even exchange parity, 
but the underlying grade-1
generators $\gamma_{\alpha}$ are now odd,
fermionic.

\subsection{Fermionic aggregates}
\label{sub:FERMIONIC}

If the Fermi-algebraic quantum theory of gravity is correct,
space-time is a singular limit of a fermionic aggregate.
In any case, the main sources of gravity are fermionic,
not bosonic, so we require
the familiar algebra of
 an aggregate $\0M E$ of fermionic entities $E$
 (soon to be events).

\begin{assert} A spin $1/2$  is a fermionic aggregate.
\end{assert}
{\bf Argument}:\hspace{2pt}
\BE
\label{eq:SPIN}
\0M\, V =  \Vee V = \Min \Fermi V= \Min \Cliff \Dup V =\Spin \Dup V
\EE
with $\Dup$ of Definition \ref{def:DUP}, \S\ref{sec:FERMI}, and
$\dag$ induced by that of $V$. 
To see the last equality in (\ref{eq:SPIN}), recall that
for Cartan 1935 and Chevalley 1954,  
$\Vee V$ is the prototype spinor space
for a neutral quadratic space isomorphic to $V\oplus V\dual$\/.
That is, the kets supporting the algebra $\Fermi V$ are  
the columns of a faithful irreducible representation of 
$\Cliff \Dup V$, hence spinors;
and are also multivectors of the Grassmann algebra $\Vee V$.
$\square$
\vskip10pt
Usually one expresses gravity using spinors of the Minkowskian tangent space. 
This clashes with canonical quantum theory. 
Spinors require
a bilinear form $\ddag$,  
canonical quantum theory a Hermitian form $\dag$. 
This discord already exists between Minkowski space-time 
and Hilbert space.
Since we now regard the usual real Minkowski space-time as a
singular limit of a quantum space-time with a Hermitian form,
the discord disappears.  
We can express flexed gravity using multivectors over 
a fermion ket space.

$\Vee V=\Min\Fermi V$ is a kind of square root of $\Fermi V$.
The dimension of $ \Dup V$ is $2n$, the dimension of 
$\Cliff \Dup V$ is $2^{2n}$,
and 
\BE
\Dim  \Min\Cliff \Dup V =\sqrt{2^{2n}}=2^n\,.
\EE

Before encountering 
the Segal homotopy strategy, I 
vacillated between using $\Cliff V$ as algebra and as ket space.
Now it is clear that the regular case 
$\Fermi V=\Cliff \Dup V$ is the full matrix algebra,
provided with intrinsic inner product,
while its singular limit $\Grass V$ 
is only a ket space,
requiring an external inner product.

\subsection{Bosonization}
\label{sub:BOSONIZATION}
It is especially easy to construct paleo-bosonic 
excitation quanta 
out of a fermionic aggregate. 
\begin{assert} A fermionic aggregate
includes a paleo-bosonic aggregate.
\end{assert}
{\bf Argument}\hspace{2pt}
The vectors $\gamma_{\alpha}$ of the Fermi algebra
are fermionic generators, and the
tensors $\gamma_{\alpha\dag\beta}$
are 
unitary-group generators and therefore
paleo-bosonic generators that
have bosonic generators as singular limit. $\square$
\vskip5pt
The even sub-algebra $C_+\subset C$ 
of Fermi statistics
can be interpreted as
the ket space of an aggregate
composed of fermion-pairs  $\gamma_{ab}$,
of even exchange parity.

History $H$
will be presented
as a fermionic aggregate  of events
$E={\mu} H$
that are in turn
fermionic aggregates of mini-events $\mu E = \mu^2 H$. 

\subsection{The cosmological number problem}
There is no sign
of basic space-time aggregates intermediate 
between nuclear sizes and cosmic.
Therefore
we must
leap from  nuclear to  cosmic  sizes
by a single fermionic quantification;
just
as classically one leaps from differentials to stellar distances
by one integration.
\begin{assert}
The number of distinguishable events 
$E=\mu H$ in the maximal system history $H$
is not greater than the dimensionality of the event ket space:
\BE
\label{eq:NUMBER}
N \le \Dim \0V\,E .
\EE
\end{assert}
{\bf Argument}\hspace{2pt}
Since $H =\0M E$, 
a basis of the ket space $\0V \,H$ 
can be composed of products 
of orthonormal kets in a basis of $\0V E$, each to some power.
The number of kets with non-zero powers is an eigenvalue of $N$.
This no greater  than the number of kets in the basis. $\square$
\vskip 10pt
Every model of event space
since Archimedes is an instance of this assertion.
Nowadays $N$ has to be large enough to account
not only for the vastness of space-time but also  all for the fields that are assigned to the event in the singular limit of classical space-time.
The successes of the continuum limit suggest that
\BE
 \0P^5 1 =2^{16}\sim 10^{5} \ll N  <\Dim \0V\, E <
\0P^6 1 =2^{2^{16}}\sim 10^{10^{10}}\/.
\EE
$N$  is
 so large that I assume that the event $E=\mu \, E$
  too is composite.
Its elements are mini-events  $X={\mu} E$, elsewhere called chronons.
By Assumption \ref{ass:BFS} of basic Fermi statistics,
\BE
\Dim\0V\,X \sim  \log_2 \Dim \0V\,E \sim 2^{16}\/.
\EE
We can reduce the event dimensionality
 if we restrict quantum theory 
to
a logarithmically small part of the universe,
leaving the rest for the meta-system including the experimenter.
For example  even if  the number of events in the history of the universe 
is $\0N U\sim 10^{20\,000}\approx 2^{60\, 000}$,
the number $\0N H$ of events in the maximum feasible quantum 
system history $H$ is no more than 
about 60\, 000,  very roughly speaking,
and  three levels
of fermionic analysis bring us to binary elements: 
$E\sim \0P^3 2= 65 \,536$\/.
\BE 
\Dim \0V E\sim \0P^3 \22\/.
\EE
Nevertheless, for an aggregate of 60~000 fermionic events to be possible, 
the generic event must have kets of 60~000 components,
and the chronon must have kets of $~16$ dimensions/
Macroscopic space-time  extension singles out 4 of the 16,
as though a condensation of
many elements $\mu^2 E=\0M\22$.

\subsection{General spin-statistics correlation}
\label{sub:GSS}
For any entity $E$ let $X(E)$ 
be the {\em exchange parity} of $E$, 
the operator of an aggregate $\0M E$
that exchanges two elements $E$ by a homotopy.
Let $W(E)$
be the (Wigner) spin parity of $E$,  
representing a continuous spatial rotation of  one element
$E$ in
the aggregate $\0M E$ through $2\pi$.
$W(E)$ has eigenvalue $+1$ for integer spin and
$-1$ for half-odd spin. 

The observed spin-statistics correlation is 
\BE
W(E)=X(E)
\EE
 for all
quanta $E$.

The quantifier $\0M$  converts a single $E$ into an entity composed of a
variable number of $E$'s described by a spinor or multivector.
The spin-statistics correlation is then the M-M correlation:

\begin{assert} [M-M correlation]
If $A=\0M B$ is a  fermionic aggregate of quanta in space-time
then its generic element is also a
fermionic aggregate:
\BE
A=\0M  B \quad \0{implies}\quad  B=\0M C
\EE
in nature.
\end{assert}
{\bf Argument}\hspace{2pt}
Since $A$ is a fermionic aggregate, $B$ must be a fermion. Then by the spin-statistics correlation, $B$ is represented by a spinor. 
A spinor is a multivector. $\square$
\vskip 10pt
 The assertion simply restates the spin-statistics correlation.
 
$\0M $ and $\Cliff$  both exponentiate the dimension:
\BE
\Dim\0M  C = \Dim  \Cliff C = 2^{\Dim C}
\EE
But their difference is essential. 
To
iterate $\Cliff$  would seriously violate the spin-statistics correlation,
To iterate $\0M=\Min\, \Cliff \Dup$ supports it:

\begin{assert}   Basic Fermi statistics (Assumption
\ref{ass:BFS})
implies the spin-statistics correlation.
\end{assert}
{\bf Argument}\hspace{2pt}
According to Assumption
\ref{ass:BFS} 
the quantifier $\0M $ is applied at every level,
and an M-M correlation exists throughout the vertical structure.
The spin-statistics correlation is merely the special
case where the middle level is that of quanta in space-time.
$\square$
\vskip 10pt
Conversely, if we accept the existence of one basic statistics,
the M-M correlation implies that 
it is the Fermi statistics.

\section{General quantum covariance}
\label{sec:GQC}

General relativity 
replaces Minkowski space as a model for event space
by a  space with much larger group:
\begin{definition}
\label{def:EINSTEINLIE}
The {\bf Einstein group} ${\0E}(\3M)$ is the group
of diffeomorphisms
$\3M\to \3M$
of the manifold $\3M$ whose points represent physical events, and whose
metrical form is  the gravitational or chronometric form
$g_{\mu\nu}(x)$.
The {\bf Einstein Lie algebra} $d\,{\0E}(\3M)$
is the Lie algebra of the Einstein group. 
\end{definition}
$d\,{\0E}(\3M)$ consists of the smooth real vector fields $X=(X^{\mu}(x)\partial_{\mu})$
on $\3M$, taken with the Lie product $X_1\x X_2:=[X_1,X_2]$.

As Einstein based general relativity on covariance under
the Einstein group, to maintain correspondence
I  base general quantum relativity on
covariance under a quantized Einstein group $\8{\0E}$,
a flex of the Einstein group ${\0E}(\3M)$.

\subsection{Non-locality}
\label{sub:NONLOCALITY}
Ultimately ${\0E}(\3M)$ is compound and therefore singular
because its elements respect 
the underlying points of the manifold, 
mapping points into points. 
This form of locality
results in invariant subgroups and algebra ideals
and must be eliminated in full quantization.

Flexing does this at an adjustable length.
Vilela space and Baugh space (like Snyder space) 
already have a non-locality with range
characterized by  a new quantum  constant of time.
Since  they
are  still without gravity,
their theories might be called special quantum relativity,
of the $D$ and $A$ series respectively.
I carry this non-locality 
into general quantum relativity.

Quantizing the Einstein group is easier
if we base 
it on an algebra instead of a manifold.
One may redefine the Einstein  group
as the
automorphism group
of the  algebra $A(\3M)$ of manifold coordinates:
\BE
{\0E}(\3M)=\0{Auto} \,A(\3M).
\EE
To quantize ${\0E}(\3M)$ I quantize the algebra
$A(\3M)$ and take the automorphism group of the result.
This drops smoothness and locality, which have no meaning
for the quantized space
and must re-emerge in the singular limit.

\subsection {Renunciation of absolute space-time}

Quantum events like those of Vilela space or Baugh space
have, in each admissible frame,
not only space-time coordinates 
but also momentum-energy  and other coordinates,
mixed by the invariance group.
This is counterintuitive
since it relativizes
the construct of absolute space-time point
that has pervaded physics since Galileo if not Aristotle.

Were Galileo and Einstein right on this point,
there would 
exist entities underlying mechanics
that have space-time coordinates but no momentum-energy coordinates:
the space-time points.
If we trace this notion back to the slate  or
the sandy beach on which Euclid  and Archimedes developed and
tested their ideas of geometry,
we see that their ``points'' were actually
bits of mineral.
They had well-defined momenta that were small because they 
and the observer are both well-coupled to each other and
to one large condensate, the Earth.
Space-time is thus an extrapolated abstraction from the 
solid-seeming Earth.
Therefore its elements too may have
momentum coordinates that have been similarly suppressed.
Quantum 
 events are merely reclaiming the energy-momentum variables
 that were abstracted from them millennia ago.
 These energy-momentum variables may have zero vacuum expec tation valiues, 
 but they will inevitably contribute to the vacuum stress tensor,
 and possibly to dark matter and the cosmological constant.

\subsection{Vilela space}

I use the following familiar structures:
\begin{definition} 
\label{def:HG}
The $2N+1$-dimensional  {\bf Heisenberg Lie algebra} 
 $d\,\0H(N)$ is defined by the relations
\BE  
[p_{\nu}, q^{\mu}]= ir, \quad [r, p_{\mu}]=0=[q^{\mu}, r]
\EE
among $2N+1$ Hermitian generators $q^{\nu}, p_{\nu},
r$.

 The {\bf Heisenberg group} ${\0H}(N)$ is
the $2N+1$ dimensional Lie group
infinitesimally generated by $d, {\0H}(N)$.

The {\bf Heisenberg algebra} $A_{\0H}(N)$
is the algebra of bounded operators
providing
an irreducible faithful representation $R_{\hbar}\0H(N)$
 with  invariant $r=i\hbar$\/.

A {\bf Heisenberg space} $S_{\0H}$
is a quantum space whose coordinate algebra 
is the Heisenberg algebra and whose 
group of allowed coordinate transformations
is the Heisenberg group.
\end{definition}

Vilela 1994 flexed the Heisenberg Lie algebra to
an orthogonal Lie algebra $d\,\0{SO}(6;s)$
with generators 
$o_{\alpha\beta}$.
 The Vilela coordinates are  
 \BEA
 \label{eq:VILELA}
 \8x^{\alpha}&=& \4X o^{\alpha \, X},\cr
\8p_{\alpha} &=&\4P o_{\alpha Y},\cr
 \8L_{\alpha \beta} &=& 
 \8x_{\alpha}\8p_{\beta}-
 \8x_{\beta}\8p_{\alpha}=\4X\4P o_{\alpha\beta},\cr
 \Fi&=&l^{-1}o_{XY}
 \EEA
 with  scale factors $\4X$ and $\4P$
having the units of length and momentum.

The quantum number $l$ is the maximum eigenvalue
of $-io_{XY}$ in the representation $R_J$.
 
The contraction to 
classical space-time  includes the limits
 \BE
 \label{eq:CONTRACTION}
\4X, \4P\to 0, \quad 
\4N, l\to \infty
 \EE
 and the freezing
 \BE
 -io_{XY}\approx l\/.
 \EE
 
\subsection{Unitarization}

$d\,\0{SO}(6;s)$ can be imbedded isomorphically
in a (pseudo-) unitary Lie algebra 
$d\, \0{SU}(6,s)$ by enlarging the underlying real vector space
$V$ with bilinear form
$g_{\alpha \beta}$ to a complex vector space $V\ox \1C$ 
 sesquilinear form 
$g_{\alpha^* \beta}$ 
with the same numerical coefficients in some frame 
$\3B$:
\BE 
g_{\alpha \beta}\buildrel{\3B}\over =  g_{\alpha^* \beta}
\EE
This process resembles the Weyl unitary trick;
Let us call it {\em unitarization}.
The generators of the orthogonal group $\0{SO}(V)$
are antisymmetric matrices
$o_{\alpha\beta}$.
Unitarization replaces these real antisymmetric matrices by matrices 
that have the same matrix elements in the frame $\3B$, 
so I will continue to designate them by $o_{\alpha\beta}$.
It
appends an equal number of 
imaginary symmetric matrices $s_{\alpha\beta}$, and a smaller number of imaginary traceless diagonal matrices 
$d_{\alpha}$,  generators $\0{SU}(V\ox \1C)$.
It is convenient to use an overcomplete set of generators subject to
$\sum_{\alpha} d_{\alpha}=0$.

Instead of the usual infinite-dimensional representation
$R_{\hbar}$ 
of 
the Heisenberg Lie algebra $d\0H(\3N)$ by differential operators,
one must choose among an infinite number of
finite-dimensional representations
$R_{J}d\0{SO}(\nu;\sigma)$ 
of the unitary Lie algebra 
$d\0{SU}(6;s)$ in $d\0{SU}(N;S)$,
labeled by the appropriate collection $J$ of quantum numbers.
$J$ determines
the dimension $N$ and signature $S$ of the representation space. 
In what follows, 
a circumflexed variable is the $R_J$ representative of the  
un-circumflexed variable.

\subsection{Baugh space}
Baugh space is then a unitarization
of Vilela space.
There is no need to rewrite the defining relations
(\ref{eq:VILELA})
but unitarization gives the symbols different meanings for Baugh space.

To imbed the  Baugh algebra 
$d\,\0{SU}(V)$, and so the Vilela algebra,  
within a Fermi algebra $\Fermi V$
(\S\ref{sec:FERMI}) 
one chooses an arbitrary orthonormal basis $\3B$
of  vectors $|n\>\in V$  and writes
$\<\iota |n\>\in \Fermi  V$ for the corresponding  creators.  
$\<\iota|: V\to \Fermi V$ is thus an operator-valued form
characterizing fermionic statistics.
$\<\iota |n\>$ can be read as ``create $n$".
The corresponding  annihilator
can be written as $\<n|\iota\>$,
``$n$-annihilate''
with $|\iota\>: V^{\4D}\to \Fermi V$
an operator-valued vector, acting from the right.
Then the Fermi representative of any $u\in \0{GL}(V)$ is
the well-known operator-valued expectation value
\BE
R_N u = \sum_{m,n}\<\iota |n\>\<n|u|m\> \<m|\iota\>=:
\<\iota|u|\iota\>\in \Fermi V\/.
\EE
We use this to imbed $\0{SU}(V)$ and $d\,\0{SU}(V)$
in $\Fermi V$\/.

The Bose Lie  algebra $\0B(V)$  is
generated by vectors $v\in V$ and $v\adj\in V^{\4D}$
with $[u, v\adj]=v\adj\bullet u$\/ and $[u,v]=0$\/.
Obviously $\0B(V)\lrar \0H(V)$:  the Bose and Heisenberg Lie algebras are isomorphic.
I apply our prior flex of $\0H(V)$ to $\0B(V)$:

To flex $\0B(V)\fro\0{SU}(V\oplus \22)
\subset\Fermi(V\oplus \22)$,
append two additional basis elements 
$\gamma_X,\gamma_Y$ orthogonal to $V$
to form the vector space $V":=V\oplus \22$\/.
Then  flex
\BEA
v\to \8v&:=&\<\iota|v\>\<X|\iota\>\in \Fermi (V")\/,\cr
v\adj\to \8v\adj_k&:=&\<\iota|Y\>\<v|\iota\>\in \Fermi (V")\/. 
\EEA
The flexed commutation relations for vectors $v,w\in V$ follow:
\BEA
[\8v,\8w]&=&0,\cr
[\8v\adj, \8w]&=& \<\iota|Y\>\<X|\iota\> \<v|w\>,\cr
[\8v\adj,\8w\adj]&=&0,
\EEA
Other variables 
vanish or freeze in the singular limit of
the usual  bosonic statistics.
This limit 
first ``freezes''
the imaginary generator $i o_{X Y}$
to a small sector of its ket space 
where $i o_{\0X \0Y}$ 
is close to its extreme eigenvalue $\4N$.
 The sector must be large enough to 
 support all the kets of the singular limit,
 yet small enough so that $io_{XY}\approx \4N$\/.
 Typically I consider neighborhoods 
containing $\0O(\sqrt{\4N})$
eigenvalues of $i o_{\0X \0Y}$ or fewer
 as $\4N\to \infty$ in
 the canonical limit.
 
\subsection{General quantum relativity group}
\label{sub:GQCA}

For purposes of quantization, let us redefine 
the general relativity group
 (diffeomorphism group) in more algebraic language:

\begin{definition}
The {\bf general relativity group} or {\bf Einstein group}
${\0E}(\3N)$ of a 
classical space-time $\3N$ (\em redefined)
is the 
 group of smooth local automorphisms of
the commutative coordinate algebra $\2A(\3N)$ of all coordinate
functions on $\3N$.
\end{definition}
Then to quantize the Einstein group it remains only to
quantize the algebra it acts on.

General relativity (or general covariance)
contrasts with special relativity in that it
 implicitly assumes that the coordinates on physical space
make up a commutative algebra.
The fully quantum correspondent is a full matrix algebra.
The Assumption \ref{ass:BFS} of 
basic Fermi statistics implies that this
is a Fermi algebra, that of the ket space of the 
mini-event $\0M\,E=:X$: $\0A\, E=\Fermi \0V\,X$\/.

\begin{assumption}[General quantum relativity] 
\label{ass:GQR}
The  event 
is a Fermi aggregate.
The generalized event coordinate is an 
operator on the ket space of the aggregate.
The general quantum relativity group  is the
automorphism group of the algebra of event coordinates.
\end{assumption}
I write $E=\0M\, X$ for the event and $G$ for the group.
$G$
must be distinguished from the automorphism group of the 
graded Lie algebra  $d\,\0{SU} (\0V \,E)\subset G$,  
a much smaller group appropriate 
to special quantum relativity.

It follows that 
the ket space of the event is $\0V E = \Vee \0V X$,
with dimensionality  $\nu:=\Dim \0V\, E= 2^{\Dim \0V\, X}$\/.
The special quantum coordinates of the event
form not an algebra but merely the Lie algebra
$d\,\0{SU}(\0V E)\subset \Fermi \0V X$\/,
imbedded in the Fermi algebra by the representation
\BE
d\,\0{SU}(\0V E)\ni u \mapsto \<\iota |u|\iota\>
\in \Fermi \0V E
\EE

The  general quantum coordinates form the algebra 
$\Fermi \0V X$.
The general quantum relativity group  $G$ is the
group of the regular elements of  $\Fermi \0V X$,
modulo its center $\1C$\/.

This brings the groups of  general relativity and quantum theory into close alignment.  
Both  are plausible contractions of one general quantum  relativity group $\0{SU}(\nu;\sigma)$ with event ket space
$\0V E = \1C(\nu;\sigma)$, a complex $\dag$ space of dimension $N$ and signature $S$.

The main sequence of contractions or singular limits   
now proposed is
\begin{quote}
General quantum space $\to$ Baugh space $ \to$ 
Heisenberg space $\to$ Minkowski space.
\end{quote}
 Classical and semiclassical general relativity lie on another line of contractions that dangles from the left-most space of this sequence.
Spinors arise as  singular limits of multivectors.
 
This puts the diffeomorphism groups of general relativity
 and the unitary groups of quantum theory into precise alignment. 
 Both are plausible 
 contractions of the general quantum relativity group,
 along different contraction paths.

By the assumption of Fermi statistics
the kets of  the physical event and the physical history are multivectors and therefore spinors. 
There is no free choice of representation, as there is for
Vilenka space, Baugh space, and paleo-bosonic statistics.
Fermi algebras have unique faithful irreducible representation.

For the fermionic event $E=\0M \, X$ 
with ket space $V=\0V E$,
the quantum history $H=\0M\,E$ has 
 ket space $\0V H = \0V\0M E = \0M V$
 and coordinate algebra $\0A H=\Fermi V$\/.

Every Fermi algebra $\Fermi V$
has a Hermitian norm
\BE
\label{eq:NORM}
\|x\|:=\frac 1N \0{Tr} \,x\adj x\/
\EE
and a quadratic form
\BE
Q(x) := \frac 1N \0{Tr} \,x^2
\EE
 This quadratic form is indefinite, as is needed for physics, 
 of signature $N(N+1)/2 -N(N-1)/2 = N$,
 the square root of its dimension.
 For example, it is a Minkowskian form of signature 2
 on the $2\x 2$ matrices

\section{General quantum kinematics}
\label{sec:GQK}
Einstein used  a scalar-valued
 quadratic form $v^{\mu'}(x)g_{\mu'\mu}(x)v^{\nu}(x)$
 on vectors to describe gravity.
 In canonical quantum gravity this form would be 
operator-valued.

\begin{assumption}  [Theory of gravity]
General quantum gravity is described by 
the  operator-valued quadratic form $v^2$,
where $v$ is a cliffor 
(Clifford element)
that has $v(x)$ as singular limit and the product 
is the Clifford product.
\end{assumption}
{Argument}\hspace{2pt}
A plausibility argument occupies
the rest of \S\ref{sec:GQK}.
\vskip10pt

The gravitational quadratic form maps vectors to scalars.
But vector fields in turn are derivations (differentiators)
on scalar fields.
Therefore in order to  quantize gravity
 I  first  quantize the scalar fields,  then the vector fields, 
 and finally gravity.
 
\subsection{Clifford ring of classical gravity}
\label{sub:RING}
For completeness I review the 
Clifford ring $C(\3M)$ of classical gravity
on a space-time manifold $\3M$,
 giving its scalars,  vectors, and
 product,
before quantizing it.

By a local ring on a manifold I mean one with local (associative) 
product and sum; that is,
the values of the sum and product at a point
are determined by the values of their arguments
at that point. 

\begin {definition}
The {\bf gravitational Clifford ring} $C(\3M)$ 
of a classical gravitational manifold
is 
a local Clifford ring 
with these properties:
\begin{enumerate}
\item The scalars  of
$C_0(\3M)\subset C(\3M)$
are the smooth
 scalar functions
$\3M\to \1R$.
\item The vectors of 
$C_1(\3M)\subset C(\3M)$
 are the vector fields, the derivations on $C(0(\3M)$.
\item  ({\em Clifford law}) The square of a vector is a scalar.
This scalar then defines
a local quadratic form in the vector, 
\BE
\|v(x)\|=g_{\mu\nu}(x)v^{\mu}(x)v^{\nu}(x)\/. 
\EE
\item This is the gravitational quadratic form of proper time.
\end{enumerate}
\end{definition}

In other words,
classical gravity is the part of the structure tensor
of the associative Clifford ring of space-time
that couples two classical vector fields to a classical scalar field;
and it determines the rest of the structure tensor.
One finds the gravitational quadratic form at a point 
by limiting the vector fields to a small
neighborhood of the point and measuring proper times.
This requires arbitrarily small clocks and rods,
so it does not work below a certain size.

$\3M$ must have  suitable global topology
(second Stiefel-Whitney class 0) 
for its gravitational
Clifford ring $C(\3M)$ to admit 
a globally defined spinor module $\Psi(\3M)$ with
$C(\3M)=\0{Endo}\,\Psi(\3M)=\Psi\ox \Psi\adj$.

\subsection{Quantized scalars}
\label{sub:SCALARS}
Recall that the algebra $A(\3N)=C_0(\3N)$ of scalars
on a space-time manifold $\3N$ is
the grade-0 part of 
the Clifford ring $C(\3N):=\Cliff(\3N)$ 
of \S\ref{sub:RING}.
It has the following properties:
\begin{enumerate}
\item   $A(\3N)$ is an associative unital commutative
real algebra.
\item $A(\3N)$ contains and is generated by
the coordinate functions $x^{\mu}$ of one frame
and the imaginary unit
$i$.
\item $A(\3N)$  is invariant under Einstein $\0E(\3N)$.
\item $A(\3N)$ is commutative.
\end{enumerate}

The third condition
incorporates the principle of equivalence.
The fourth condition, commutativity, implies 
that $A(\3N)$ is the algebra of functions
on the state space of some classical object,
here
the generic event of $\3N$.
The quantum correspondents would seem to be:
\begin{assumption}
The ring $\8A$
of scalar functions 
in general quantum relativity
regarded as quantum coordinates of the quantum event,
has the following properties
relative to the Clifford  and Fermi algebra
$\8C=\Cliff \Dup V = \Fermi V\supset \8A$:
\begin{enumerate}
\item $\8A$ is a full matrix algebra.
\item $\8A$ includes
the quantum coordinate functions $\8x^{\mu}$
and imaginary $\Fi$\/.
\item $\8A$  is invariant under $G$.
\item $\8A$ is minimal in the above respects.
\end{enumerate}
\end{assumption}
 By conditions 2 and 3, $\8A\supset \8C_+$, 
the even-grade subalgebra of $\8C$, do thst necessarily
\BE 
\frac 12 \Dim \8C_+\le \Dim \8A\le \Dim \8C.
\EE
I satisfy these condition by assuming $\8A=\8C$\/: 
The quantum scalars form the entire 
Clifford-Fermi algebra.
 This obviously satisfies all the conditions but minimality,
 which I conjecture to hold.
 
The coordinate contraction 
$\8x^{\mu}\to \7x^{\mu}, \Fi\to i\hbar$ induces
an algebra contraction $\8{A}\to A(\3N)$\/.

The grade-2 cliffors of the form
 $\gamma_{\mu X}, \gamma_{XY}\in\8C$
 represent  the action of
 infinitesimal orthogonal unitary transformations
 $o_{\mu X}, o_{XY}: V\to V$ 
upon multivectors of $\Vee V$.
 The special quantum  coordinate $\8x^{\mu}$ is
\BE
\8x^{\mu}= \4X \gamma^{\mu X}:\8C\to \8C.
\EE

The  coordinates commute in the contraction limit
$\4X\to 0,\, \4N\to \infty$\/.
One may perform  the contraction 
 to  the canonical limit
by changing two generators of the Clifford-Fermi algebra
$\8C$
from
$\gamma_{X}, \gamma_{Y}$ to $\4X\gamma_{X},
 \4P \gamma_{Y}$,
 freezing
 $\gamma_{XY}^2$,
and taking the limit 
 \ref{eq:CONTRACTION}
 of the structure tensor.
The freezing consists in
 restricting the ket space
 to a subspace in which $(\gamma_{XY})^2$ 
 is relatively near its extreme value $-(\4N)^2$.
 Then
 the general quantum operator
 \BE
 \Fi\hbar:=\frac{\gamma_{XY}}{\4N}\/.
 \EE
reduces to the usual quantum imaginary $i\hbar$\/. 
 
This gives $\8A$ a large new center in the limit of classical space-time,
the ring of complex scalars $\Cliff_{0}(\3N)\subset \8C$.

\subsection{Quantized vectors}
\label{sub:VECTORS}
The defining property of the contracted 
vector fields $\gamma_{\mu}(x)
\in \Cliff_{1}(\3N)$
is that they are the derivations of the ring of scalars
$\Cliff_{0}(\3N)$ over the field $\1R$:
\BE
\Cliff_{1}(\3N):=\0D \Cliff_{0	}(\3N)\/.
\EE
They form a Lie ring over the algebra of scalars
$\Cliff_{0}(\3N)$.
I therefore define:
\begin{definition} 
The {\bf general quantum vectors}  are elements of the
Lie algebra of
derivations $\Delta\8A$ of the general quantum scalar algebra 
$\8A= \8C$.
\end{definition}
That is, the same Clifford-Fermi
algebra $\8C$
 that describes quantum scalars by its associative structure
describes quantum vectors by its commutator structure.
This vector construction is 
invariant under the quantized Einstein group
and contracts to the usual concept
in the space-time limit.

\subsection{General quantized gravity}
By the $\0M \0M $ hypothesis
(\S\ref{sub:UNISTATISTICS}),
the ket spaces of the quantum history $H$,  
the quantum event $E={\mu}  H$, 
and the quantum mini-event  $X={\mu} ^2 H$ 
are multivector and spinor spaces:
\BE
\0V\, H=\Vee  \0V\, {\mu}  H=\Vee ^2 \0V\, {\mu}^2 H.
\EE  
In the general quantum theory 
I take
the gravity form to be the quadratic form of the Clifford product
of the event algebra $\8C=\Fermi \0V E=\Cliff \Dup V$.
That is,  the general quantum operator-valued
bilinear form $\8g_{\circ\circ}$
of gravity is
 the tensor representing the bilinear form 
 $\8g(u)=u^2=u^2$\/,
the  square
of the cliffors that contract to vector fields in the limit 
of classical space-time.
Let $A, B, \dots$ be collective indices of the event
space.
Then
\BE
\label{eq:g}
\8g_{\{A|B\}}:=
frac 12 \{\gamma_{A}, \gamma_{B}\}\/.
\EE

The skew-symmetric part of the same  product 
 defines the infinitesimal generator 
$\gamma_{[A|B]}\in d\8{\3E}=d\0{Iso}(\8C)$
of the general quantum  Einstein group:
\BE
\gamma_{[A|B]}\:=
\frac 12 [\gamma_{A},\gamma_{B}]=-\gamma_{[B|A]}
\EE
In this general quantum kinematics 
the gravitational field is not a function on a space-time manifold.
The c space-time manifold is
an organization or condensation of the quantum event space.

\section{General quantum dynamics}
\label{sec:GQD}
I may still assume that the dynamical history multivector 
is  an  exponential
\BE
\Omega =e^{iA/\hbar}\in \Vee \0V\, E
\EE
where now $A$ is an action multivector.

Here I make another tactical decision.
Rather than 
quantizethe Hilbert action,
which is not quantum general covariant,
I seek quantum general covariant second-order
dynamical equations.

What corresponds to general covariance  is
\begin{assumption} [General quantum covariance] 
The action
$A$ is invariant
under the  group of transformations 
$\0V\,H\to \0V\,H$
induced by the general quantum covariance group 
$G:
\0V E\to \0V E$
through the Fermi quantification relation 
$H=\0M  E$.
\end{assumption}
\begin{assert}{The action $A$ is a 
polynomial in the Casimir operators of the unitary group
$G$.}
\end{assert}
{\bf Argument}\hspace{5pt}Clear.$\square$

Einstein assumed that the dynamical law was.
expressed by second-order wave equations.
This led to an action that is second order in $p$. 
but  not in $x$\/. 
If  we
 imitate him too closely we would  
 break the $o_{XY}$ symmetry
 between $x$ and $p$, 
violating general quantum covariance.
 \begin{assumption}  
The length and momentum scales of experiments so far, 
$X_{\0{Exp}}$ and $P_{\0{Exp}}$, are related to
the quantum inits by
\BE
\4X\ll X_{\0{Exp}}, \quad
\4P\gg P_{\0{Exp}}
\EE
\end{assumption}  
This would explain why we have been led to a singular theory with
 $x$-locality  and  $p$-non-locality.
Now we must give up the $x$-locality, 
restoring $x\lrar p$ symmetry.
Then I can suppose, following Einstein and Hilbert,  that 
\begin{assumption}
The action is second-order in the generators of the  group 
$\0{SU}(\0V E)$.
\end{assumption}
This singles out the Casimir invariant of the  unitary group of 
the quantum event $E$
as represented in the unitary group of the quantum history $H$.
If $u_A$ is a basis for $d\,\0{SU}(E)$
and $\8u_{A}=\<\iota|u_A|\iota\>\in \0A\, H$
 is the representation
of $u_A$ in $\0A\, H$
then the Killing form is
\BE
\label{eq:K}
K_{AB}=\0{Tr} \, \Delta \8u_A \, \Delta \8u_B\/,
\EE
and the action is the Casimir invariant  with a  multiplier
coupling constant $m$:
\BE
\label{eq:A}
A=m  K^{AB}\8u_A\8u_B\/.
\EE

\section{Discussion}

\label{sec:DISCUSSION}

\subsection{Roads to the quantum event}
The photon, the quantum of the electromagnetic oscillator,
 had to be verified several times to be generally accepted.
 The quantum event
 and  sub-event
  will require at least as much scrutiny.
 On the one hand, they
attack even deeper continuity assumptions
 than the photon did;
 on the other hand, physicists are have more experience with
such reformation processes today.
The three main
roads to the photon were:
\begin{enumerate}
\item Regularization. 
 Planck first introduced the quantum constant $h$
to eliminate the infinite classical heat capacity of  
electromagnetic cavities.
  
\item One-photon observations. 
Einstein recognized  $h$ as the action quantum of a single photon
and estimated it independently from the photoelectric effect.
Compton confirmed the photon and estimated $h$ yet again
by bouncing photons one at a time off free electrons.

\item Quantization.  
Dirac deduced the photon from the canonical commutation relations for the electromagnetic field.
\end{enumerate}

It is harder to see
one event   than 
one photon.
Space-time is stiff 
 while cavity photons form an ideal gas,
so events in a cavity are strongly coupled while
the photons in the same cavity are weakly coupled.
 To predict  
quantum effects that are presently observable
we must formulate a theory of  
many strongly-coupled events
before we ever  isolate a single one.
We did not have this problem with the photon.

Since we cannot reach
 the
quantum event
  by road 2 yet,
I have approached it
  by roads 1
and 3.

\subsection{The kinematics of general quantum gravity}
I have argued that the
 macroscopic Clifford ring $C(\3M)$ of classical gravity
on a space-time manifold $\3M$
is a singular limit of
an underlying Clifford algebra $\8C$  of 
general quantum gravity;
that the macroscopic Lie product of classical gravity
is a singular limit of the  Lie algebra
of the even-grade cliffors of $\8C$,
with the commutator as Lie product;
that the classical gravitational field is a
singular limit of the
general quantum  operator-valued quadratic form
defined by the Clifford product of $\8C$.

The Einstein group
respects the Lie product
but not the  ring product.
The quantized Einstein group is therefore 
the special unitary  group $\0{SU}(E)$
of  the event ket space.
The correspondence 
from 
$\0{SU}(E)$ to  the Einstein group is
the main tool  of this paper.
It leads to the
general quantum gravitational kinematics of (\ref{eq:g})
and the dynamics of (\ref{eq:A}).

General relativity  renounced 
the idea that space-time coordinates
have immediate metrical meaning
but retained the absolute points themselves.
General quantum relativity 
renounces absolute space-time points for
a unified quantum concept of space-time,
angular momentum, and momentum-energy.
The methods apply to gauge theories in general.

Dynamics is defined here not by  
a one-parameter group of unitary operators
generated by the Hamiltonian
but by a ket 
defining a probability amplitude for quantum histories,
corresponding to the classical action.

\subsection{Nonlocality}
To take this extension of relativity
 seriously we must overcome the enormous
apparent difference between space-time coordinates and 
momenta in our current experience.
The difference boils down to the usual
 assumption
that Nature  can
make jumps in $p$ but not in $x$\/;  or
that fields are diagonal in $x$ but not in $p$.
 
In general quantum relativity I must suppose that this is
 a broken $x\leftrightarrow p$
 symmetry. 
When we restore the broken $x\leftrightarrow p$ symmetry we 
lose
locality in $x$ as well as $p$.
Infinitesimal locality is not even defined for general quantum variables, which have discrete spectra.
Quantum theory already permits us to interchange $x$ and $p$ by Fourier
transformation.
In Vilela space, $x$ and $p$
are interchanged by the operator $o_{XY}$; 
in Baugh space by $o_{XY}$ and $s_{XY}$.
Since breaking $o_{XY}$ invariance might break
$i$ invariance, it is good that we have 
$s_{XY}$ invariance to break instead:
another amenity of the $A$ series.

The spectral gap $\4X$ in $x$ is also a measure of 
the non-local jumps in $x$.
The difference in locality
that we currently see
reflects
 a real difference in the ranges $\4X$ and $\4P$
of 
the quanta of $x$ and  $p$
on the scale of present quantum experiments.

\section{Acknowledgments}

I gratefully acknowledge discussions 
with James Baugh, Eric Carlen,  
Andrei Galiautdinov, Alex Kuzmich, 
Frank Schroek, Mohsen Shiri-Garakani, 
Heinrich Saller, and
Julius Wess, who shared Heisenberg's
early thought on quantum space-time.

\section{Appendix: Clifford algebra}
\label{sec:CLIFFORD}

I use a Clifford construct that covers the following cases:
\begin{enumerate}
\item The Clifford algebra of Chevalley,  with an arbitrary field of coefficients 
and an arbitrary quadratic space. 
\item The Clifford algebra of the usual quantum theory of spin,  with
complex field and Minkowskian quadratic space.
\item The Clifford ring of general relativity,  with coefficients in the  ring of scalar fields, which is not a field, and  quadratic space composed of the vector fields.
\item The Clifford algebra of every Fermi algebra,  with  complex field
and complex neutral quadratic space.
\item
The Clifford algebra of classical set theory,  with scalars in the binary field 2,
and  vectors in  the second power set of an arbitrary finite universe of discourse. Its  Clifford product has $\gamma^2=1=\emptyset$ and
is a XOR operation; its Clifford sum has
$\gamma+\gamma=0$= the undefined and
is the addition of Boole as extended by Pierce.
\end{enumerate}
Let $C_1$ be a $\dag$ module called the vectors. 
Call its commutative unital 
ring of coefficients $C_0$ the scalars.
\begin{definition}
A {\bf Clifford ring} over $C_1$
is a
ring  $C\supset C_1 $  
generated by
 $C_1$ that
obeys Clifford's Law in the form 
\begin{quote}
{\em The square of any vector is a scalar .}
\end{quote}
The {\em Clifford form} is
the quadratic form $N: C_1\to C_0, \;
v\mapsto v^2=:\|v\|$   on $C_1$,
{\bf Cliffors} are			
elements of $C$\/.
A {\bf Clifford algebra} is a Clifford ring whose scalars 
form a field.  
\end{definition}

$C$ defines a bilinear inner product for any vectors $u, v\in C_1$: 
\BE
u\cdot v = \frac 12 \{u,v\}\/.
\EE
I assume here that this bilinear form is regular.

A Clifford ring with product $u\sqcup v$ 
defines a  Grassmann ring with the same elements
with Grassmann product $a\vee b$ such that for vectors
$u, v\in C_1$
\BE
u\sqcup v = u\cdot v + u \vee v\/.
\EE
The Grassmann ring has a well-known $\2N$-valued grade.
This grades the Clifford ring too.
Insofar as its Grassmann algebra
can be interpreted as the algebra of a fermionic aggregate,
so can  a Clifford algebra.

The Clifford norm  is invariant under
automorphisms of the Clifford algebra. 
Its signature is determined by the signature and dimension
of the quadratic space $C_1$.

\begin{definition} 
The {\bf spinor space} $\Spin V$ over a quadratic space $V$ 
consists of the columns
of (a faithful irreducible matrix representation of) the 
Clifford algebra $\Cliff V$ \cite{BUDINICH}:
\BE
\Spin V= \Min  \Cliff V
\EE
\end{definition}

\section{Appendix:Fermi algebra}
\label{sec:FERMI}
The Fermi algebra of Fermi-Dirac statistics is a 
Clifford algebra
on the $A$ series, close in spirit to Cartan's
original construction of spinors for the $D$.

\begin{definition} Let $K$ be a ring with a quadratic
norm $\|k\|=k\adj k$. Let $V$ be a module over  $K$
(usually a vector space over a field). Then
$V\adj$ designates the {\bf dual module} 
of $K$-linear mappings $V\to K$.
$\dag: V\to V\adj$ designates a singled-out non-singular anti-linear involutory map
$V\to V\adj$, 
$V\adj\to V$,  $K\to K$,
agreeing with the given $\dag$ on $K$.
A {\bf $\dag$ module} is a module endowed with a $\dag$.
A $\dag$ algebra is a $\dag$ space  with an algebra product
of which $\dag$ is an involutory anti-automorphism:
$(ab)\adj=b\adj a \adj$\/.
$\Dim V$ is the dimension of $V$ over $K$.
\end{definition}

Let $V$ be a  $\dag$ space over $\1C$, and let
$\4D V= V^{\4D}$ be the dual space
of linear maps $V\to \1C$.
Let $f\bullet x := f(x)$ represent application of
a map $f$ to its argument $x$.

\begin{definition}

\noindent$\bullet$ $\Fermi V$, the {\bf Fermi algebra} over $V$,
 is 
the  $\dag$ algebra 
generated by $V, V\dual  \subset \Fermi V$
subject to the Fermi-Dirac relations: 
\BE
 \forall v, w\in V
 \quad \Bar \quad
  (v+w\adj)^2=\|v+w\adj\|= w\adj \bullet v+
v^\adj\bullet w.
\EE
\vskip5pt
\noindent$\bullet$ $\forall f\in \Fermi   V\quad\Bar\quad 
\|f\|:=\0{Re}\, f^2\/. $
\vskip5pt
The space $\0M V$ of {\bf multivectors}
is the $\dag$ space  
$\Vee V\subset \Fermi V$ 
with the $\dag: \0M V\lrar \0M V\dual$ induced
by the $\dag$ on $V$\/.
\noindent $A$ is the endomorphism algebra of $V$
and $\0M A$ is the endomorphism algebra of $MV$.
\noindent {\bf Multi-operators} are operators on multivectors;
elements of $\0M A$.
\noindent$\bullet$ For any unitary transformation $u:V\to V$, 
$\0M u:\0M V\to \0M V$  is the induced automorphism
and $\0S u\in V\vee V\dual\subset Fermi V$ is any element
generating $\0M u$ as an inner automorphism:
$\0M u = \0S u \,v\, (\0S u)^{-1}$.

$\square$
\end{definition}

\noindent{\em Standard interpretation}: 
\vskip5pt
\noindent$\bullet$ $V$ is the ket space of a fermionic quantum entity $E$, representing input channels for the entity.
A $\dag$ on $V$ induces one on $\Fermi V$.
\vskip5pt
 \noindent$\bullet$ $\0M V$ is the ket space of a many-$E$ aggregate $\0M E$.
 \vskip5pt
  \noindent$\bullet$ $\Fermi   V$ is the coordinate algebra of the aggregate $\0M E$.
\vskip5pt
\noindent$\bullet$ $\|f\|$ is the transition probability amplitude for the process
represented by  $f^2 = f\circ f$. $\square$

\vskip5pt
$\vdash:\quad  \Dim \0M V=2^{\Dim V}\/.$

A Fermi algebra is a Clifford algebra:
\begin{definition} \label{def:DUP}The {\bf duplication space} of $V$ is
\BE 
\Dup V:=V\oplus V\dual
\EE
with quadratic form $\ddag$ defined 
for any  $ v\in V, w\in V\dual, (v,w) \in \Dup V$
by
\BE
\|(v, w)\|= (v,w)^{\ddag}(v,w):=w(v)\/.
\EE
\end{definition}
The duplication space was introduced by Saller 2006a as the {\em quantum
space} of the system.  
$\Dup V$ has both a neutral quadratic form and a hermitian form induced by that of $V$.
\begin{assert} Every Fermi algebra is a Clifford algebra.
\end{assert}
{\bf Argument}\hspace{2pt}
\BE
\Fermi V= \Cliff \Dup V \quad\square
\EE

The $\dag$ on the multivector (spinor) space $\0M V$ is required to be invariant
under $\0{SU}(V)$. 
This singles out the $\dag$ on
$\Vee V$ 
induced by the $\dag$ on $V$, with the property that 
\BE
\forall u, w\in 
\0M V\quad \Bar \quad 
(u \vee w)\adj=w\adj vee u\adj.
\EE
This is positive definite.
The indefinite metrics that we need in physics
cannot come from this $\dag$.
They may arise from the neutral quadratic form on $\Dup V$.

The first natural Fermi multivector spaces $\0M^L \1C$:
\BEA
\0M^0 \1C&=& \1C,\\
\0M^1 \1C&=& 2\1C,\\
\0M^2\1C&=&4\1C,\\
\0M^3\1C&=&16\1C,\\
&\vdots &\\
\0M^L\1C &=& \0P^L 1\/.
\EEA
$\0M2$ is a four-dimensional space with Minkowskian quadratic form.


\end{document}